\documentstyle[prb,aps,epsf,twocolumn]{revtex}
\begin{document}
\draft

\title{Polarizabilities of Germanium Clusters}
\author{Jinlan Wang$^1$, Mingli Yang$^2$, Guanghou Wang$^1$, Jijun Zhao$^3$}
\address{$^1${\it National Laboratory of Solid State Microstructures and Department}\\
{\it of physics, Nanjin University, Nanjing 210093, P.R. China}\\
$^2${\it Department of Chemistry, Nanjin University, Nanjing 210093, P.R.}\\
China\\
$^3${\it Department of Physics and Astronomy, University of North Carolina }%
\\
{\it at Chapel Hill, Chapel Hill, North Carolina 27599-3255}}
\maketitle

\begin{abstract}
Polarizabilities of Ge$_n$ clusters with 2 to 25 atoms are calculated using
coupled-perturbation Hartree-Fock (CPHF) and finite field (FF) method within
density functional theory. The polarizabilities of the Ge$_n$ clusters
increase rapidly in the size range of 2 to 5 atoms and then fluctuate around
the bulk value. The polarizabilities are sensitively dependent on the
cluster geometries and electronic structures. The large HOMO-LUMO gap may
lead to the small polarizability. As compared with the compact structure and
diamond structure, the prolate cluster structure corresponds to a larger
polarizability. 

\end{abstract}
\pacs{36.40.-c, 36.40.Cg, 61.46.+w, 71.24.+q}

In the past two decades, the structural and electronic properties of
semiconductor clusters have been extensively studied because of their
fundamental interest and potential application in nanoelectronics \cite
{atom1,mass1,mass2,mass3,frag,phonoi,pes1,pes2,pes3,ion,ion1,Rata,4,5,Ale,Wang,zhao,Menon,Ralasubramanian,Lanza,Deutsch,Edet,Ogut,Li,Lu}. The small semiconductor clusters are well understood up to 10 atoms. But
our knowledge for larger clusters is still quite limited. The polarizability
is one of the most important quantities of the clusters, which can yield the
static dielectric constant in the bulk limit through Classius-Mosotti
relation. On the other hand, the polarizability can provide some information
on the bonding and geometrical features of the clusters. Thus,
comprehensively understanding of the polarizabilities from theoretical
calculations is important in cluster science. The jellium model \cite
{jellium,jellium1} was successfully applied to study the polarizability of
large metallic clusters. But for semiconductor clusters, the bonding and
geometrical effects are not incorporated in the jellium model.
Alternatively, the {\em ab initio} calculations based on quantum chemistry
methods are needed. There were only few previous {\em ab initio}
calculations on the polarizabilities of the clusters \cite
{Vasiliev,fue,Jackson,Vasiliev1,Deng,prak}. Especially, for the germanium
clusters, there is only one attempt to calculate the polarizabilities and
the cluster size is quite limited \cite{Vasiliev}.

In our previous studies \cite{jin}, the geometries of Ge$_n$ clusters ($%
n=2-25$) have been obtained by density functional DMol calculations \cite
{dmol} incorporated with a genetic algorithm \cite{ge1,ge2,ge3}. It was
found that the Ge$_n$ clusters follow a prolate growth pattern with 
$n\geq 13$. The stacked layered structure and the compact structure
compete with each other in intermediate size range. Based on the previously
optimized low-energy structures, in this paper, we will study the
polarizabilities of these clusters using the analytically
coupled-perturbation Hartree-Fock (CPHF) and numerically finite field (FF)
methods within density functional theory. We aim to explore the size
dependence of the polarizability, the influence of the atomic and electronic
structures on the polarizabilities of the Ge$_n$ cluster.

It is well known that electron correlation plays a primary role in
determining molecular polarizabilities. DFT treatment has been proven to
make significant improvement to Hartree-Fock results in molecular
polarizabilities and hyperpolarizabilities calculations\cite{Mat,per}.
Within DFT framework, B3LYP functional considers the hybrid between the
Hartree-Fock exchange and Kohn-Sham orbitals\cite{b3lyp,becke,lee}, while
LANL2DZ basis set \cite{lanl2dz} can give a good description of the bonding
and geometrical features of heavy atoms. Thus, B3LYP/LANL2DZ scheme is
expected to well describe the cluster polarizabilities at acceptable
computational cost. Here, all these calculations are performed at
B3LYP/LANL2DZ level by using Gaussian98 package \cite{gaussian}.

The energy in an external electric field can be expanded as 
\begin{equation}
E(F)=E(0)-\mu _iF_i-\frac 12\alpha _{ij}F_iF_j-\cdot \cdot \cdot
\end{equation}
where $E(0)$ is the energy without the external field, $F_i$ are components
of the applied field. The dipole moment $\mu _i$ and the polarizability $%
\alpha _{ij}$ are defined as:

\begin{equation}
\mu _i=-\left( \frac{\partial E}{\partial F_i}\right)
\end{equation}
and

\begin{equation}
\alpha _{ij}=-\left( \frac{\partial ^2E}{\partial F_i\partial F_j}\right) 
\begin{array}{llllll}
&  &  &  &  & (i,j=x,y,z)
\end{array}
\end{equation}
Within Kurtz's finite field method\cite{finite}, the dipole moment and
polarizability in a uniform field, can be derived from the follow equations.

\begin{equation}
\mu _iF_i=-\frac 23\left[ E(F_i)-E(-F_i)\right] -\frac 1{12}\left[
E(2F_i)-E(-2F_i)\right]
\end{equation}

\begin{equation}
\alpha _{ii}F_i=\frac 52E(0)-\frac 43\left[ E(F_i)+E(-F_i)\right] +\frac 1{12%
}\left[ E(2F_i)-E(-2F_i)\right]
\end{equation}
To calculate of the dipole $\mu $ and the polarizability $\alpha $, at least
13 self-consistent field (SCF) runs are necessary with the field strengths $%
\pm F_i$ and $\pm 2F_i$ ($i=x,y,z$). One of the most severe problems in
finite-field method is the choice of an appropriate field strength\cite
{field}. Sim {\em et al.} \cite{sim} have assessed the numerical accuracy
against different field values and concluded that stable linear and
nonlinear polarizabilities can be obtained when $F$ equals $0.001\sim 0.005$
a.u. In this work, the external field is added along $x,y,z$ with the
magnitude 0.001 a.u and a tighter SCF convergence criterion of 10$^{-9}$ is
adopted.

The measured data in experiments are usually the average polarizabilities,
which can be obtained by

\begin{equation}
<\alpha >=\frac 13(\alpha _{xx}+\alpha _{yy}+\alpha _{zz})
\end{equation}

The optimized structures for Ge$_{11-25}$ clusters at B3LYP/LANL2DZ level
are shown in Fig.1. Similar equilibrium configurations are found by DMol and
Gaussian98. Therefore, the details for the structures will not be discussed
here.

\begin{figure}
\vspace{-0.15in}
\centerline{
\epsfxsize=3.5in \epsfbox{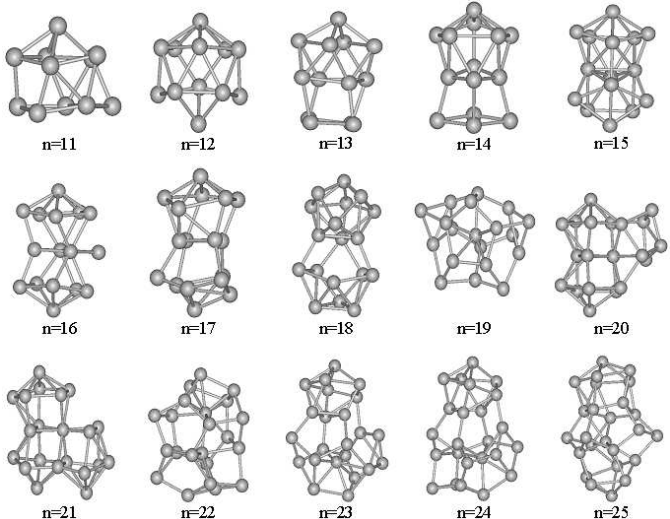}
}
\caption{Lowest energy structures for Ge$_n$ ($n=11-25$) clusters.}
\end{figure}

\begin{table}[tbp]
Table I. The dipole moments $\mu$(Debyes) , polarizabilities $\alpha($\AA $%
^3/$atom) of Si$_{2-10}$ clusters:$\alpha^a$: present results, $\alpha^{b}$:
Ref.\cite{Vasiliev},$\alpha^c$: Ref.\cite{Deng}, $\alpha^d$: Ref.\cite
{schafer}, (in bracket):Ref.\cite{Vasiliev}.
\par
\begin{center}
\begin{tabular}{cccccc}
$n$ & $\mu$ & $\alpha^a$ & $\alpha^b$ & $\alpha^c$ & $\alpha^d$ \\ \hline
2 & 0(0) & 4.97 & 6.29 &  &  \\ 
3 & 0.32(0.33) & 4.56 & 5.22 &  &  \\ 
4 & 0(0) & 4.48 & 5.07 &  &  \\ 
5 & 0(0) & 4.87 & 4.81 &  &  \\ 
6 & 0.01(0.0) & 4.62 & 4.46 &  &  \\ 
7 & 0(0) & 4.60 & 4.37 &  &  \\ 
8 & 0.23(0.0) & 4.83 & 4.52 &  &  \\ 
9 & 0.17(0.36) & 4.73 & 4.38 & 4.46 & 3.0 \\ 
10 & 0.29(0.69) & 4.55 & 4.31 & 4.65 & 5.50
\end{tabular}
\end{center}
\end{table}

To check the validity of current method, we firstly calculate the dipole
polarizability of small silicon clusters. Table I compares our calculations
with previous theoretical and experimental results. The current theoretical
dipole moments for the clusters Si$_{2-7}$ are in agreement with Vasiliev 
{\em et al}\cite{Vasiliev}. But our results for the dipole moment of Si$%
_{8-10}$ clusters are different due to the substantial difference in
geometrical configurations. The present geometric structures for Si clusters
are consistent with those obtained by Shvartsburg {\em et al.} and Li {\em %
et al.} \cite{Ale,Li}. Our calculated polarizabilities seem to be
overestimated as compared with Vasiliev {\em et al}\cite{Vasiliev}, but
consistent with Deng {\em et al}\cite{Deng}. Moreover, the present
polarizabilities are sensitively dependent on the cluster size and oscillate
with the cluster size. While the polarizabilities in Vasiliev's work tend to
decrease with the increase of the size. Experiments show that the
polarizabilities fluctuate with the cluster size.

\begin{table}[tbp]
Table II. The dipole moments $\mu$ , polarizabilities $\alpha($\AA $^3/$
atom), binding energies $E_b$ (eV), HOMO-LUMO gaps $\Delta$ (eV) of Ge$_n$
clusters: $\mu_{\text{CPHF}}$(Debyes): CPHF method; $\mu_{\text{FF}}$
(a.u.): FF method; in bracket: Ref.\cite{Vasiliev}.
\par
\begin{center}
\begin{tabular}{cccccc}
$n$ & $\mu_{\text{CPHF}}$ (PRL) & $\mu_{\text{FF}}$ & $\alpha$ & $E_b$ & $%
\Delta$ \\ \hline
2 & 0(0) & 0 & 4.10(6.67) & 1.69 & 0.31 \\ 
3 & 0.61(0.43) & 0.241 & 5.07(5.89) & 2.84 & 1.32 \\ 
4 & 0(0) & 0 & 5.14(5.45) & 3.23 & 1.14 \\ 
5 & 0(0) & 0 & 5.52(5.15) & 3.32 & 1.48 \\ 
6 & 0.15(0.14) & 0 & 5.36(4.88) & 3.41 & 1.39 \\ 
7 & 0(0) & 0 & 5.27(4.70) & 3.50 & 1.36 \\ 
8 & 0.55(0) & 0.22 & 5.47(4.99) & 3.44 & 1.10 \\ 
9 & 0.12(0.28) & 0.05 & 5.39(4.74) & 3.48 & 1.22 \\ 
10 & 0.56(0.68) & 0.22 & 5.17(4.66) & 3.59 & 1.31 \\ 
11 & 1.35 & 0.53 & 5.24 & 3.53 & 0.99 \\ 
12 & 1.58 & 0.62 & 5.33 & 3.49 & 1.05 \\ 
13 & 0.82 & 0.32 & 5.45 & 3.57 & 0.98 \\ 
14 & 1.30 & 0.51 & 5.41 & 3.61 & 1.20 \\ 
15 & 0.18 & 0.07 & 5.44 & 3.57 & 0.80 \\ 
16 & 0.74 & 0.29 & 5.39 & 3.61 & 1.17 \\ 
17 & 1.11 & 0.24 & 5.52 & 3.59 & 0.94 \\ 
18 & 0.61 & 2.75 & 5.61 & 3.58 & 0.90 \\ 
19 & 0.46 & 0.18 & 5.41 & 3.63 & 0.70 \\ 
20 & 1.06 & 0.46 & 5.51 & 3.61 & 0.89 \\ 
21 & 1.70 & 0.67 & 5.54 & 3.62 & 0.94 \\ 
22 & 0.51 & 0.20 & 5.45 & 3.63 & 0.81 \\ 
23 & 2.89 & 1.14 & 5.62 & 3.62 & 0.82 \\ 
24 & 2.52 & 0.99 & 5.69 & 3.64 & 0.62 \\ 
25 & 0.52 & 0.21 & 5.63 & 3.61 & 0.70
\end{tabular}
\end{center}
\end{table}

Table II gives the dipole moments, average polarizabilities, binding
energies and HOMO-LUMO gap as functions of the cluster size. The dipole
moments calculated with CPHF method and FF method are compared and agree
well with each other. The dipole moment reflects the symmetry of the
geometrical structure: the smaller dipole moment corresponds to the higher
symmetry. The dipole moments for the clusters with $n=2,4,5,7$ are nearly
zero, corresponding to the high symmetry in these clusters. For Ge$_4$, the
lowest energy structure is D$_{2h}$ rhombus, while they are trigonal
bipyramid with D$_{3h}$ and pentagonal bipyramid (D$_{5h}$) for Ge$_5$ and Ge%
$_7$. In the case of Ge$_8$ , Ge$_9$ and Ge$_{10}$, capped pentagonal
bipyramid, tricapped trigonal prism and tetracapped trigonal prism have
favorable energy, respectively. The dipole moments and the polarizabilities
of small Ge clusters are compared with Vasiliev {\em et al}.\cite{Vasiliev}
and our calculations seem to be a little overestimated.

\begin{figure}
\vspace{0.65in}
\centerline{
\epsfxsize=3.0in \epsfbox{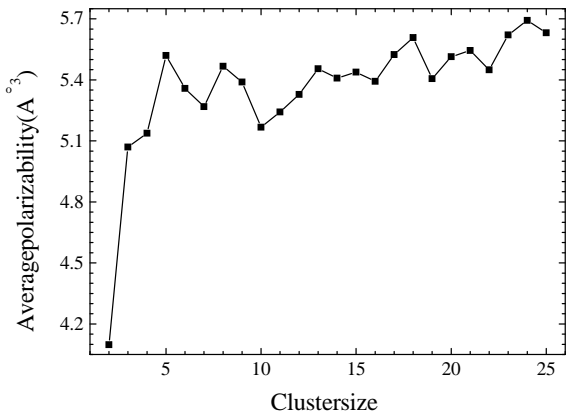}
}
\vspace{-0.75in}
\caption{The polarizabilities of Ge clusters are plotted as a function of the cluster size for Ge$_n$.}
\end{figure}
Fig.2 gives the polarizabilities of Ge$_n$ clusters as a function of cluster
size. The polarizabilities of small clusters increase rapidly with cluster
size in the size range of 2 to 5 atoms. Then, the polarizabilities fluctuate
around $\alpha =5.4\AA ^3/$atom. In the light of Clausius-Mossotti relation,

\begin{equation}
\alpha =\frac{\epsilon (\omega )-1}{\epsilon (\omega )+2}r^3
\end{equation}

the polarizability of the bulk Ge is $4.5$\AA $^3/$atom. The present average
polarizability of the clusters is larger than that of the bulk. Previous
studies have shown that the prolate structure contributes a large
polarizability\cite{Jackson,guan}. The prolate configuration is preferred
for medium-sized Ge clusters, which occupies larger distortion and lower
symmetry in comparison with diamond structure. Thus, the polarizability of
the clusters is larger than that of the bulk. It is worthy to note that the
present results is quite different from Vasilier's\cite{Vasiliev}. Compared
with their method, the present scheme incorporate the electron correlation
in the calculation of the polarizability of small clusters.

We further discuss the relationship between the polarizability and the
electronic structures of clusters. Fig.3 plots the polarizabilities as a
function of the HOMO-LUMO (highest occupied molecular orbital (HOMO) and
lowest occupied molecular orbital (LUMO)) energy gaps. As shown in Fig.3,
the large polarizability generally corresponds to the small HOMO-LUMO gap.
The clusters with $n=8,11,13,15$ have relatively smaller polarizabilities in
comparison with their neighboring size. For example, the HOMO-LUMO gap for Ge%
$_8$ is 1.10eV and the polarizability is just 5.17 \AA $^3/$atom. The
polarizabilities decrease with the increase of the HOMO-LUMO gap except for
the clusters Ge$_2$, Ge$_5$ and Ge$_{19-22}$. This can be easily
rationalized using the two-level model\cite{twolevel1,twolevel2}, 

\begin{equation}
\alpha \sim \frac{\mu _t^2}{\Delta _t}
\end{equation}

where $\mu _t^2$ is the transition dipole moment from the ground state to
the first dipole-allowed excited state and $\Delta _t$ the corresponding
transition energy. Approximately, $\Delta _t$ can be replaced with HOMO-LUMO
energy gap $\Delta $. From this model, $\alpha $ increases with decreasing $%
\Delta $ , consistent with our calculated trend for most Ge clusters.
However, $\alpha $ is not a simply inverse proportion relation to $\Delta $,
since the two quantities are dependent each other. Small $\Delta $ tends to
generate large $\mu _t^2$. In addition, $\mu _t^2$ depends on some other
characters, such as selection rule.

\begin{figure}
\vspace{0.65in}
\centerline{
\epsfxsize=3.0in \epsfbox{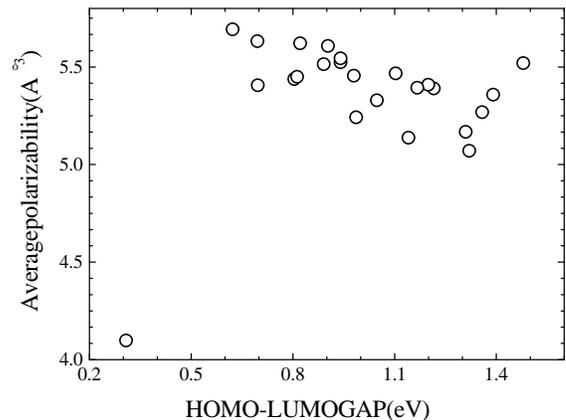}
}
\vspace{-0.75in}
\caption{Polarizabilities of Ge clusters are plotted against the HOMO-LUMO gap.}
\end{figure}

However, the polarizability of the cluster is not only dependent on
HOMO-LUMO gap but also closely related to geometrical characteristics. For
example, the HOMO-LUMO gap of Ge$_{18}$ is larger than that of Ge$_{19}$,
while the polarizability of the former is larger than that of the latter.
This can be attributed to their structural difference. For Ge$_{18}$, the
optimized structure is two interpenetrate pentagonal connected with a
bicapped square antiprism Ge$_{10}$ subunit, while a more compact cage-like
geometry is favorable to Ge$_{19}$. The compact structures have relatively
less and shorter bonds, which leads to the valence electrons binding
tighter. Thus, a smaller volume is obtained in the compact structure, which
causes a smaller polarizability for Ge$_{19}$. Similarly, the different
polarizabilities behavior of the clusters with $n=19-22$ can be explained in
the light of their respectively geometrical characteristics. For Ge$_{20}$
and Ge$_{21}$, their optimized structures are stacked layer configurations
and have the comparable volume, which leads to their almost same
polarizabilities. In the case of Ge$_{19}$ and Ge$_{22}$, since our
calculated lowest energy structures are both near-spherical compact
structures, they also have the tantamount polarizability.

In conclusion, we have calculated dipole moments and polarizabilities of Ge$%
_n$($n=2-25$) clusters with both CPHF and FF approaches under B3LYP/LANL2DZ
scheme. The main results are summarized as following. (1) The Ge clusters
with $n=2,4,5,7$ have relatively higher symmetry and the dipole moments are
nearly zero. (2) The polarizabilities of small clusters increase rapidly in
the range of 2 to 5 and fluctuate around 5.4 $\AA ^3/$atom. Moreover, the
polarizabilities of the clusters with $n=8,11,13,15$ are larger than the
neighboring ones. (3) The polarizabilities are closely related to the
HOMO-LUMO gaps and the geometrical configurations. The larger the HOMO-LUMO
gap, the smaller the polarizability of the Ge clusters. The prolate
structures corresponds to relatively large polarizabilities in comparison
with the compact structures.

\ \\ 

This work is financially supported by the National Natural Science Foundation of China(No.29890210 and 10023001).

\end{document}